\documentclass{IOS-Book-Article}

\usepackage{mathptmx}
\usepackage{soul}\setuldepth{article}
\usepackage{graphicx}
\graphicspath{ {./images/} }

\def\hb{\hbox to 11.5 cm{}}

\begin{document}
\pagenumbering{arabic}
\pagestyle{headings}
\def\thepage{}

\begin{frontmatter}              

\title{The games we play: critical complexity improves machine learning
}

\markboth{}{June 2022\hb}

\author[A]{\fnms{Abeba Birhane}} and 
\author[B]{\fnms{David J. T. Sumpter}}

\runningauthor{Birhane and Sumpter}
\address[A]{Mozilla Foundation \& School of Computer Science, University College Dublin, Ireland}
\address[B]{Department of Information Technology, Uppsala University, Uppsala, Sweden}

\begin{abstract}
When mathematical modelling is applied to capture a complex system, multiple models are often created that characterize different aspects of that system. 
Often, a model at one level will produce a prediction which is contradictory at another level but both models are accepted because they are both useful. Rather than aiming to build a single unified model of a complex system, the modeller acknowledges the infinity of ways of capturing the system of interest, while offering their own specific insight. We refer to this pragmatic applied approach to complex systems --- one which acknowledges that they  are incompressible, dynamic, nonlinear, historical, contextual, and value-laden --- as Open Machine Learning (Open ML). In this paper we define Open ML and contrast it with some of the grand narratives of ML of two forms: 1) Closed ML, ML which emphasizes learning with minimal human input (e.g. Google's AlphaZero) and 2) Partially Open ML, ML which is used to parameterize existing models. To achieve this, we use theories of \textit{critical complexity} to both evaluate these grand narratives and contrast them with the Open ML approach. Specifically, we deconstruct grand ML `theories' by identifying thirteen 'games' played in the ML community. These games lend false legitimacy to models, contribute to over-promise and hype about the capabilities of artificial intelligence, reduce wider participation in the subject, lead to models that exacerbate inequality and cause discrimination and ultimately stifle creativity in research. We argue that best practice in ML should be more consistent with critical complexity perspectives than with rationalist, grand narratives.

\end{abstract}
\end{frontmatter}
\markboth{June 2022\hb}{June 2022\hb}

\section{Introduction}
\label{sec:introduction}

Over the past decade, Machine Learning (ML) models have come to exert significant power and influence in science and society. From predicting biological phenomena~\cite{cruz2006applications,camacho2018next}, human behaviour~\cite{rastrollo2020analyzing,salganik2020measuring,dalipi2018mooc} to social patterns~\cite{gilpin2020learning}, ML is now a widespread approach to modelling complex adaptive phenomena. 
The idea is that complex problems --- from drug design to criminal sentencing --- can be solved by obtaining large amounts of training data relevant to the problem and then using this data to find an algorithmic 
solution. 
In the process, complex issues which are intrinsically social, cultural, historical, and open ended~\cite{gershenson2013living,froese2019problem,juarrero2000dynamics} are treated as issues that can be neatly formulated into a closed question, fully captured in data, and 
``solved'' with huge volumes of data and algorithmic models.  
In order to provide a precise understanding about the ways in which ML is used in applications, we define three approaches: Closed, Partially Open, and Open ML as follows. 
\begin{description}
\item [Closed ML] Under this approach (often referred to as a black box approach), data from a system is fed in to a neural network or similar model. There is minimum or `no'~\footnote{No human intervention refers to the way a model operates. To put this in context, no model is entirely closed, in the sense that it exists independent of humans that set it up, a wealth of previous knowledge it leans on, the scientific and cultural discourse that makes such work relevant, and the environmental and mineral resources required to create and run it.} human intervention, i.e. there is a closed loop where the ML is set up and then `learns by itself'. In the process, an open, dynamic, complex phenomena is closed by the data used to train the model. 

\item [Partially Open  ML] Here, a domain expert has designed a mathematical model of a system. ML or AI methods are then used to parameterize and/or fine-tune this model. There is an interaction here between traditional modelling and ML. Interpretability is considered important, in order for the human to incorporate input into 
the machine. This approach is sometimes referred to as human-in-the-loop. In this case, the domain expert, by taking a certain phenomena and designing a model for it, partially closes an open system. 

\item [Open ML] This approach recognises that complex systems are open-ended and can never be fully captured by data or a single model. Under this approach, experts with in-depth knowledge of a subject matter use some  statistics/ML methods on data to complement that knowledge. Various different models (mathematical, visual and verbal) are constructed. Human values play a central role in this process. A variety of, sometimes contradictory and ideally modest, models of the system are created, each capturing a snap-shot of the system.

\end{description}

Our definition of Closed and Partially Open  ML is inspired by what Chantry et al.~\cite{chantry2021opportunities} called \textit{Hard} and \textit{Medium AI}, when looking at applications to weather predictions. 
All models close complex systems by artificially setting boundaries, stabilizing and freezing dynamic phenomena, and formulating them in terms of a fixed formal model. The terms Closed and Partially Open outline the extent of this closure. 
Ultimately, no model is fully closed when we consider various environmental, ecological, material resources, and continual human input necessarily required to build and maintain models. This includes teams of engineers that setup and closely monitor 
the model, the wealth of previous research that it is built on, as well as the compute power and massive amounts of data necessary to run the model. These are all necessary but often taken for granted preconditions for any model creation.

One current state-of-the-art example of  Closed ML can be found in training computers to play board games. For example, DeepMind's AlphaZero has mastered board games such as Go and Chess, as well as computer games like Space Invaders and Breakout from training conducted on pixel images~\cite{mnih2015human,silver2018general}. More complex games, such as StarCraft, can be learned using these methods, with some additional model information~\cite{vinyals2019grandmaster} (a step toward Partially Open ML). 

Partially Open  ML approaches have the potential to produce semi-automated insight into complex physical and biological systems~\cite{stevens2020ai}, such as protein-folding~\cite{noe2020machine} and improving nuclear fusion~\cite{degrave2022magnetic}. The boundary around where the systems which can be modelled by Closed ML ends and Partially Open  ML begins is contested. 
On the one hand there are suggestions that experimental procedures in some areas of physics can be run by a Closed ML~\cite{melnikov2018active}. On the other hand, Closed ML could not solve many important challenges arising from the Covid-19 pandemic for example, despite overconfident and misleading claims 
that it could~\cite{bachtiger2020machine,roberts2021common}. Chantry et al.~\cite{chantry2021opportunities} find that Closed ML performs better than Partially Open  ML at nowcasting (predicting the weather in the coming hours using data from arrays of sensors). Partially Open  ML, which includes models of atmospheric and other physical processes parameterised by ML, performs better on predictions made over a couple of days or weeks. Closed ML, which can only learn from the data it has seen and becomes unstable when extrapolating, would likely prove dangerously inaccurate if used to, for example, predict long-term climate change.

While our current work touches on the Closed/Partially Open  divide, our main focus is on the increasing number of suggested applications of Closed and Partially Open ML in predicting non-determinable, dynamic, inherently social, and contentious phenomena, such as criminal behaviour~\cite{bowyer2020criminality,keles2021cautionary,vorhees2016has}, gender~\cite{barlas2021see,scheuerman2019computers}, sexual orientation~\cite{rincon2021speaking}, trustworthiness~\cite{spanton2022measuring}, dishonesty~\cite{kamalov2021machine}, political leanings~\cite{sumpter2018outnumbered} and emotional states~\cite{stark2021physiognomic}. 
Despite the lack of scientific groundings, and with a robust body of work illustrating troubling eugenic and physiognomic roots~\cite{spanton2022measuring,y2017physiognomy,stark2021physiognomic,belden2020eugenics}, using machine learning for contentious applications is becoming normalized and widely funded.  
Affect recognition, according to Crawford, is now predicted to be an industry worth more than seventeen billion US dollars~\cite{crawford2021atlas}.

Complex phenomena, such as criminal behaviour, emerge from, are embedded in, and are entangled with social, historical, cultural, and contextual factors~\cite{spanton2022measuring,bowyer2020criminality,birhane2021impossibility} where what constitutes a crime often aligns with societal power dynamics. This means that complex behaviour is neither something that can solely be placed on individual actors (their physical appearances, body movements, or facial expressions), nor something that can be clearly defined and captured in data or a model~\cite{stark2021physiognomic}. Subsequently, these ML approaches to, for example, ``criminality prediction'' spring from a misconception that the concept of ``criminality'' can be defined unambiguously, frozen in time, and closed off. 

The rest of this paper is structured as follows. Section~\ref{sec:practice} presents practical examples of the Open ML approach, using football and biological systems as examples. Section~\ref{sec:what_is_complex} 
details what complex phenomena are, consistent with the practice in Section~\ref{sec:practice}. Specifically, we take an approach known as \textit{critical complexity}, 
adapted from works of Paul Cilliers~\cite{cilliers2016critical,cilliers2005complexity} and Alicia Juarrero~\cite{juarrero1999dynamicsbook}, which emphasizes the incompressible, nonlinear, historical, contextual, value-laden and open nature of complex systems. 
In Section~\ref{sec:games_we_play} we look closely at the `games' ML researchers play when theorising around the Closed and Partially Open approaches. The use of the term 'game' here originates with Wittgenstein~\cite{wittgenstein1953philosophical} and, later 
used by Lyotard in his book \textit{The Postmodern Condition}~\cite{lyotard1984postmodern}. Games are not entirely frivolous and the activities of applying Closed ML can provide valuable insight in some limited circumstances. However,
we argue that games involve denying complexity, and forcing Closed and Partially Open thinking on to systems that are best suited to 
an Open approach. In Section~\ref{sec:conclusion}, we argue that the `games' can, and do, lead to over-promise and over-hype about the capabilities of Artificial Intelligence (AI) systems and result is negative consequences on minoritized individuals and communities when 
these oversimplified models are applied into the social world. 

\section{The Open ML approach in practice}
\label{sec:practice}

There is 
no definitive methodology for Open ML modelling of complex systems, just a set of plural practices. In this section we 
focus on one particular application area: modelling the game of football. Team sports are more complex compared to board 
games, for example. They involve social, physical, tactical, and mental aspects. Team sports are however less complex than other 
systems such as human societies, 
financial systems, or human brains. 
Modelling the game of football, 
thus allows us to understand some of the challenges involved in modelling open systems, while still dealing with an application of (somewhat) limited scope.

\begin{figure}[t]
\includegraphics[width=11.5cm]{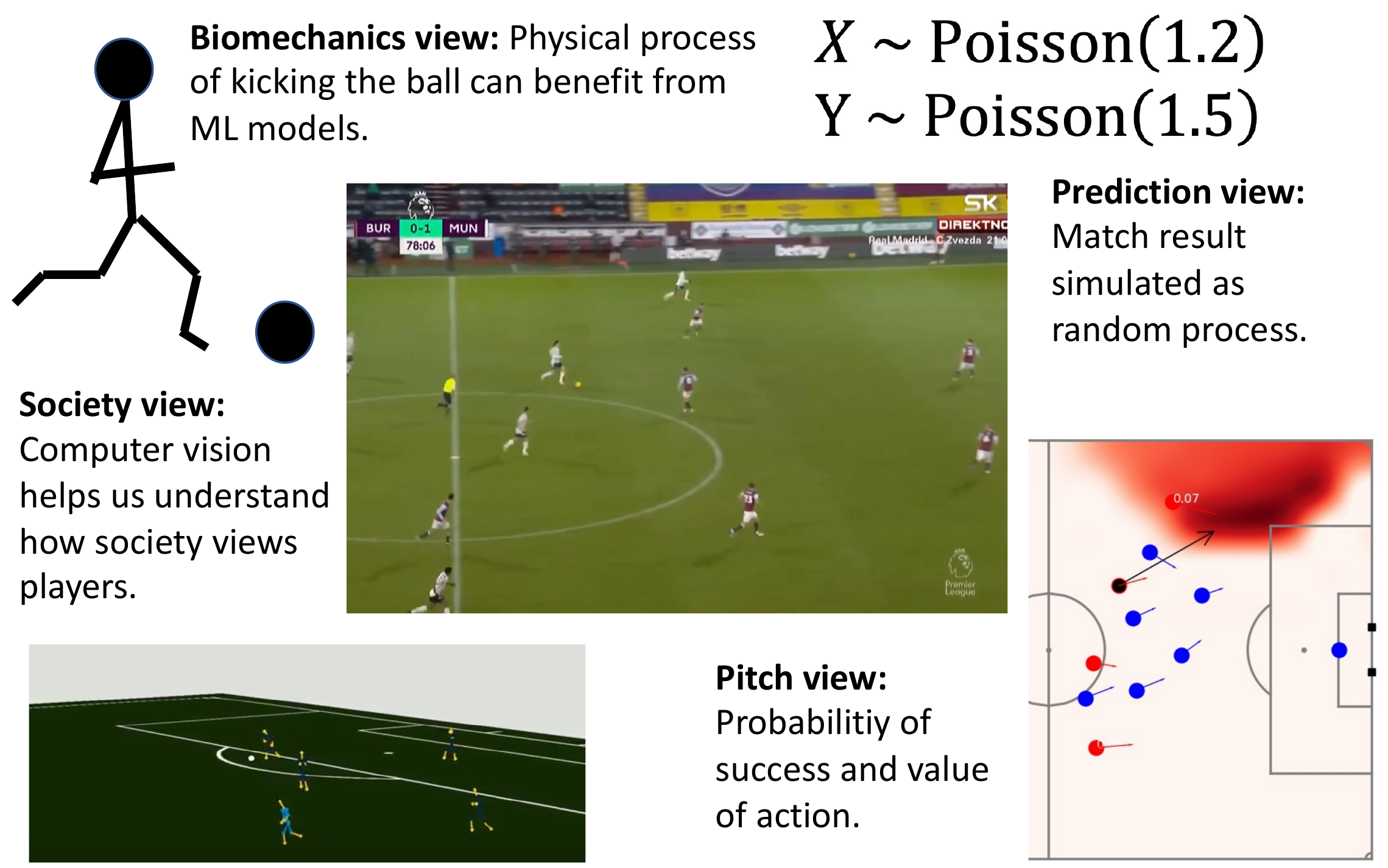}
\centering
\caption{Under the Open ML approach, the game of football can be modelled in many different ways. Here we illustrate: the prediction view (top right) simulates the game as a Poisson process; the pitch view uses models to evaluate impact of actions (bottom right); the society view uses the game to understand society at large (bottom left) (figure from \cite{Gregory2021Pace}); and the bio-mechanics view studies physical processes (top left). \label{football}} 

\end{figure}

A widely used model for predicting the outcome of a football match is Poisson regression~\cite{dixon1997modelling}. The central idea is that goals in the match 
are independent, occurring at a rate which depends on the relative quality of the teams and which can be estimated using regression methods.
This model is used by professional gamblers and bookmakers, since it outperforms betting strategies of the customers of the bookmakers (see e.g.~\cite{spann2009sports}). It is possible to include more factors, including events during the match, for example, in a neural network to improve predictions, giving a \textit{prediction} view of the game.

The prediction view is of little use to the players, who will have some sense of the strength of their opponents, and thus whether or not their team is likely to win, but can't be helped by a model (ML or otherwise) which sets probabilities to the outcome. Those playing the game want to understand specific details of their opponents' and their teammates' play which they can exploit during the match. Models that provide these insights can be found, with help of ML, through concepts such as pass probability and pass values, which (using historical data) evaluate the quality of actions~\cite{fernandez2019decomposing,sumpter2016soccermatics}. 

There are many other levels and dimensions to football, as Figure~\ref{football} shows. For example, the \textit{bio-mechanics} view looks at the body kinematics of players~\cite{ibrahim2019kinematic}. One example of the \textit{societal} view is statistical analysis of refereeing to reveal discrimination 
in decisions made~\cite{gallo2013punishing}. Another is the use of computer vision to investigate how sports commentators use words, such as `pace' and `power', when describing players with non-white backgrounds while words such as `hard work', 'effort' and 'mental skill' are used to describe white players. For example,~\cite{Gregory2021Pace} looked at how commentators described events on the pitch when they could and couldn't identify the ethnicity of the players.

Closed and Partially Open  ML models can be, and are used in the approach outlined above, in the sense that regression, neural networks, and other methods are used to fit data. But their usage is secondary to finding different views of the sport, taken from different perspectives. Finding a view is sometimes referred to in ML as feature selection. But this terminology places the ML model as primary and the features as secondary. The problem with framing this process as feature selection is that it gives the model itself an aura of neutrality to which subjectively chosen features are added. In fact, the open-ended process of model building is always a necessarily value-laden endeavour. The Open ML approach, which we emphasize, places the ML model as a tool for fitting data, once we have found the view we are interested in. Open ML, then, is about finding a useful view for a certain problem, and combining the views to get an overall understanding of the system. The usefulness of the view subsequently cannot be entirely divorced from the modeller's objectives, motivations, and perspectives.  
 
The multiple views approach is also adopted when, for example, ant pheromone trails are modelled in terms of cycles of ant activity, formation and topology of the spatial patterns of trail networks, evolution of co-operation and chemical properties of the trails~\cite{sumpter2010collective}. Further examples are found in modelling the growth of tumours, genetic networks and ecological systems. Multiple views are also a prerequisite for modelling (more complex) human social systems~\cite{helbing2010pluralistic}. In adopting an Open ML approach, we simultaneously engage many different frameworks and views of a system, each designed to answer a different sub-question. We take different snapshots of the system and then use each of them to construct a bigger picture of the system. The more snapshots we include, the more complete the bigger picture. ML might help find the sharpest focus of one particular snapshot, but it can not tell us what is a good, overall picture.

\section{What is a complex system?}
\label{sec:what_is_complex}

Sports are just one example of a complex system, which encompass phenomena ranging from physical, biochemical and biological to behavioural and social. 
Human brains, human behaviour, the financial market and society are notable examples of complex systems. A complex system necessarily consists of a large number of components, where stochastic, non-linear interactions give rise to emergent behaviour. 
The whole system cannot be fully understood from analysing individual parts. Neither is it possible to trace a neat cause-effect chain of emergent behaviour.  

These broad points about complex systems are widely recognised by mathematical modellers as well as the broader complexity science tradition which mainly comes from the 
field of physics (a la Santa Fe tradition)~\cite{mitchell2009complexity,bar1998dynamics,holland2000emergence}. 
While sharing core points with complexity science grounded in these schools of thought, our characterization of Open ML is informed by \textit{critical complexity}, building on the work of Paul Cilliers~\cite{cilliers2002complexity} and Alicia Juarreo~\cite{juarrero1999dynamicsbook}. At its core, critical complexity, emphasizes 
that complex systems are open-ended with no  
`natural' or `objectively' given boundary around where a system ends and the environment/another system begins. The observer draws the boundaries of what is included in/excluded from a given complex system, where there exist 
an infinite number of views to 
modelling a complex system. For example, the different views of football (prediction, pitch, bio-mechanical, social, for instance) came from the objectives of the stakeholders including bookmakers, coaches, players, and sociologists. Thus, when adopting an Open ML approach, our choice of what to include in a model is always made with an awareness of multiple perspectives and contexts.

Moreover, these views and perspectives are never static and change over time. The system itself changes form --- the rules change in football, animal societies change through evolutionary time, societal norms change (consider cultural and legal changes around homosexuality in the West over the past few decades) --- which means that our view has to change. The historical evolution of the system is also a potential view of the system and can be modelled. However, it is an answer to specific questions such as `how have the rules of football changed?' `why have ants evolved to follow chemical trails?' `how have attitudes towards homosexuality altered?' which are themselves open to change. 
It is impossible to close a complex system once and for all --- for this would mean that the system has come to a halt --- and imply a single `true' way of looking at it, or that there are $x$ `true' ways of looking at it. In order to model (or say something meaningful about) a system we have to draw boundaries and partially close it. However, there are infinite different ways of closing a system in the form of a model. It is this infinite number of views, some of which are more useful than others, which allows the Open ML view of modelling complexity to be self-consistent. In short, a complex system can never be closed once and for ever, but we can continue to find new and useful ways of looking at it.

The Open ML approach uses established modelling practices, while admitting that we have no value-free way of knowing whether what we have included in a model is relevant or whether what we have omitted is indeed irrelevant. While there can be, for example, a relatively clear relationship between a particular model and data (e.g. the distribution of goals are Poisson; it is more difficult to score a goal when the player is 80m from the goalmouth than 10m from the goalmouth), there can be no ``objective'' description of why someone kicking the ball between the goalposts or probability of scoring are important. 
The openness of a complex system is preserved by admitting the context-dependent choices we make when we close the system using a model, while the methods within the closed model use well-established statistical and empirical practices.

The openness of complex systems also means that they exhibit behaviour that is not reproducible in all ways. Non-linear interactions and dynamical processes can create emergent properties that are not the simple sum of components of a system. Models can help us untangle some of these non-linear relationships. However, tracing a clear causal chain all the way from the smallest, elementary components of a system, up to the largest, amalgamated description of a system is impossible given the nature of complex systems. 
Human actions cannot be explained in the same way as billiard balls~\cite{juarrero2000dynamics}. Specifically, explanations over many different levels of a system are not possible. The point here, which Anderson called `more is different', is that human psychology, for example, cannot be derived from the physics of elementary particles~\cite{anderson1972more}. As Polanyi puts it, `life transcends physics and chemistry'~\cite{juarrero2013downward}

A complex system never emerges from a historical vacuum. Its historical background partly constitutes the system's current behaviour and trajectory. According to Juarrero ``Since we carry our histories on our backs, we can never begin from scratch, either personally or as societies.''~\cite{juarrero2000dynamics}. In modelling a complex system, therefore, understanding its history is a crucial element. 
Complex systems are, by their nature, incompressible, meaning that there is no accurate representation of the system that is simpler than the system itself~\cite{cilliers2002complexity}. Building models is thus a way of capturing important (for someone, for some given purpose) elements/processes of a system, but  there is no single perfect or accurate representation or model for a given complex system. 
Accurate or perfect representation entails capturing a system (its dynamic and non-linear interactions and emergent properties) in its entirety without leaving something out. This is, in principle, impossible as closing the entire system means bringing it to a halt. The best and simplest representation of a complex system is the system itself~\cite{cilliers2002complexity}.
Each model we build provides us with a 
snapshot of that system. Since each snapshot is incomplete, they can be contradictory. Just like it is a contradiction to tell the 22 players that the outcome of the match they are about to take part in will be determined by a Poisson process.

The Open ML approach is consistent with a widely held rejection, in 
critical complexity, of various forms of positivist thinking. In a similar way that Wittgenstein argues that we need to analyse the language games we play in order to avoid falling in to philosophical traps~\cite{wittgenstein2010philosophical}, we now turn our attention to mathematical games, in the form of models, researchers play, which stand opposite to the Open ML approach. 
We look at the games researchers play, either explicitly or implicitly, in which they ignore the open nature of complex systems. We then deconstruct these games in order to avoid being misled by them.

\section{The games we play}
\label{sec:games_we_play}

In this section we look at some of the ways in which ML research treats various phenomena with a 
Partially Open or Closed ML approach, 
where 
using a Open ML or no ML at all would be more appropriate. We argue that the common underlying reason for misplaced optimism about closed forms of ML is a failure to see the complexity of the systems which it aims to model, which is then  compensated by a tendency to play games which disguise that complexity.
The core objective 
of this section is, through the lens of critical complexity, to aid  
ML researchers, and others doing mathematical modelling, identify the type of theoretical game they 
might be playing.

\textbf{The bitter lesson game:} Under Sutton's hypothesis~\cite{sutton2019bitter}, Closed ML eventually out-competes Partially Open ML, as computing power and data handling capacity increases. This may hold true for very particular applications, such as playing board games and nowcasting of the weather. 
Indeed, Closed ML  succeeds on exactly those tasks where it is easy to draw a boundary, e.g. board games have a set of relatively simple rules and the task of weather prediction involves integrating information contained in fixed atmospheric measurement devices over a very short time scale. However, there is no evidence for a bitter lesson in complex psychological or social systems. Nor is there, given the nature of these complex systems (outlined in Section~\ref{sec:what_is_complex}) 
any way forward for a closed approach. For complex systems that exist in a web of relations
embedded in social, cultural and normative contexts, it is not clear where the boundaries of the system lie, thus making it impossible to clearly delineate 
information that needs to be included or excluded in a Closed ML approach. At the very least a Partially Open  ML is needed. For complex adaptive systems to succumb to the Closed ML approach would require stripping complexity away from the phenomena and reducing them to simplistic caricatures (as is done when learning board games). 

\textbf{The map and the territory game:} Since complex systems are open systems (see Section~\ref{sec:what_is_complex}), it is impossible to perform an exhaustive mapping, formalizing, and automating of the physical, psychological/mental, and social world. Incompressible complex systems can never be captured in a single model and understood or finalized once and for all. Yet, a common game in AI and ML research is to mistakenly equate a model of a certain complex phenomenon for the phenomenon itself. A key example of this is the \textit{`Reward is enough'} thesis by Silver et al.~\cite{silver2021reward}. In this work, the authors hypothesize ``[that] reward is enough to drive behaviour that exhibits abilities studied in natural and artificial intelligence, including knowledge, learning, perception, social intelligence, language, generalization and imitation”. Through visual analogies, the authors juxtapose five systems: the board game of Go (on which reinforcement learning performs well), a robotic agent (in a computational simulation), a physical robot, a squirrel (which they interchange with human behaviour, natural agents, and animals in general) and Artificial General Intelligence (AGI). The authors commit a fallacy by moving 
from Go, to a simulated agent (which they equate to a physical robot), to biological and social agents, to AGI (a woolly and ill-defined notion) as if one can be mapped onto another. If complexity is a continuum, these systems can be viewed as increasing in degrees of complexity, from Go (simple and closed) to general intelligence (complex, open, and to a large extent hypothetical). The authors thus present  a way of thinking that seems trivially true about a closed system, but has little or no bearing on real-world open systems.

\textbf{The generalization game:} Generalization is one of the most highly desired attributes in current ML research. 
Examining underlying values of ML research through analysis of most cited papers from premier conferences in ML (ICML and NeurIPS) showed that 78\% of the top cited papers uplifted `generalization' as an important attribute~\cite{birhane2021values}. Its usage is often overloaded. For example, it is used to describe difference between train/test performance discrepancy, how well a model works on additional datasets, adding more parameters to a model and in the context of transfering a model learnt on one dataset to an application on another dataset. Within ML research, `generalization' is widely used when more precise terms such as out-of-sample, would better suffice. These usages occur against a background of the idea of a ``general AI'', which is the idea of an abstract, free-floating system that is not limited by specific use-cases. This free-standing system can supposedly be applied regardless of domain, time, or context. 

From the perspective of complexity, the very idea of creating a free-floating generalizable model or dataset that supposedly captures a complex phenomenon fully, is futile, given we can only capture a snapshot or part of a moving target. Therefore, while the specific usages (such as `out-of-sample') may be valid from a technical point of view, the way in which the word is used in ML turns it into a game. By improving out-of-sample performance, for example, a model is said to be more general, but in reality it is not general in the wider sense of the word. The inclusion of the additional dataset (on which the model is said to generalize) is simply changing the size or shape of the part of the true complex system the model encloses. Benchmark datasets themselves are inherently specific, contextualized, and finite, thus claims of general model capabilities on such benchmarks render claims of generalization meaningless~\cite{raji2021ai}. From a complexity perspective generalizability implies a model, a theory, or a method can work across different contexts, histories, and backgrounds. Contrary to this, complexity science tells us that context, history and background are crucial elements and parts of a complex system.

\textbf{The perplexity game:}
This game is seen most clearly in the evaluation of large language models~\cite{vajapeyam2014understanding,abid2021large}. Perplexity and cross-entropy are both measures of the difference between two probability distributions. For language models, these are used to measure performance, by summing the log of the probability (according to a model) of the next word in a test sequence. The game here involves relating improvements in these measures to improvements in the model. For example, Kaplan and colleagues at OpenAI emphasize scaling laws relating the number of parameters in a model and the size of the dataset to the cross-entropy error in the test set~\cite{kaplan2020scaling}. The implication here is that the bigger the model, the better it will perform and that a Closed ML approach can be applied to human language. This line of thinking exemplifies 
playing a game where form is mistaken for meaning. As Shannon wrote in his original paper on language and entropy: ``Frequently the messages have meaning; that is they refer to or are correlated according to some system with certain physical or conceptual entities. These semantic aspects of communication are irrelevant to the engineering problem.''~\cite{shannon1948mathematical}. Shannon’s original idea of entropy makes it clear that what is being measured is form and not meaning. Cross-entropy and perplexity evaluate ML models in terms of what they optimise, namely selecting a next word/string of words out of a given list, and not 
the meaning of the words which a language model produces.

\textbf{The stochastic parrot game:} In their ``On the Dangers of Stochastic Parrots: Can Language Models Be Too Big?\includegraphics[scale=0.08]{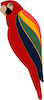}'' paper, Bender et al.~\cite{bender2021dangers} identify another game played when applying large neural networks  and the Closed ML approach to language learning: that large language models are more or less defined by the data which they are fed. All that is actually `learnt' by a Closed ML is an efficient representation of the dataset. This explains why cross-entropy decreases with parameters: the larger the model, the more efficiently data is stored. Large language models pick up and parrot patterns `learned' in datasets without any actual understanding of language.

Language is an open system that transcends the individual user. Like other complex systems, language is a dynamic living thing that is inextricably linked with cultures, histories and contexts. 
Language, therefore, is not a system that can succumb to formal rules 
but is a relational phenomenon of meaning and sense-making~\cite{di2018linguistic,cilliers2005complexity}. The meaning of language lives among the language speakers, in not only what is being said, but also in what is not being said and the way it is said. 
The important characteristics of a complex system are destroyed when it is taken apart or attempts to close it down are made. Closed ML approaches --- and, ironically given the name, OpenAI's GPT-3 is one such closed approach ––– reduce language down to form. 
Careful manipulation shows that there is no actual language understanding taking place in large language models~\cite{bender2021dangers}. Indeed,
probing state-of-the-art language models, such as GPT-2, with simple questions that require common sense understanding of everyday situations, shows these models are frequently incoherent~\cite{marcus2020next}. Understanding language remains one of AI's stumbling blocks. This, Mitchell argues, is because language is inherently ambiguous, deeply context dependent, and ``assumes a great deal of background knowledge common to the communicating parties''~\cite{mitchell2021ai}.

\textbf{The underspecification game:} One recent paper, by 40 researchers, mostly from Google, critiqued modern ML methods for being underspecified~\cite{d2020underspecification}. The authors showed that many different models, with different underlying assumptions, fit the same dataset. Such underspecification is considered as problematic, because it means the unique `correct' model cannot be identified.
Under an Open ML approach, however, the framing of the underspecification game (as described in~\cite{d2020underspecification}) can cause more problems than it solves. We expect more than one model to capture any system. Indeed, we expect lots of models to be supported by any particular piece of data. For example, models of an epidemic (the first example in~\cite{d2020underspecification}) might range from a deterministic Susceptible-Infectives-Removed (SIR) model, a stochastic model incorporating space, a full agent-based model of spread through the inhabitants of a city, a purely statistical analysis of previous epidemics, the knowledge of an experienced epidemiologist, the knowledge of another experienced epidemiologist, the wisdom of a crowd of experts and a betting market. All of these views might find some support in the existing data, but this is not necessarily  an issue in need of a technical solution. New information (such as a new variant, a better understanding of symptoms etc.) appears all the time. Thus, having a range of different, underspecified models is desirable.

\textbf{The model selection game:} While the under-specification game is about documenting ways in which different models can make the same prediction, the model selection game claims there is a method (usually in the form of Bayesian reasoning~\cite{wasserman2000bayesian}) for gradually uncovering a single correct model of a system. The idea is that by finding the model which has the largest probability given the data, i.e. which maximises $P(M|D)$, we get closer to the `true' model of a system~\cite{jaynes1988does}. Model selection implicitly accepts that the aim of modelling is to find the best single view of a system. From the perspective of a truly complex system, model selection can be a useful tool for parameter selection of any particular model, but $P(M|D)$ can not be interpreted as a probability that a model predicts the data in a wider sense, because the only full description of a complex system is the description of the system itself, which is inexhaustible. Thus Bayesian model selection is not a universal rule for these systems. A similar game is found in theories in cognitive science, more specifically 
the theory known as the \textit{free energy principle}~\cite{friston2006free}. In this theory, Friston et al. incorrectly and without basis in observation, imposes a Bayesian principle on living, cognitive 
systems~\cite{aguilera2021particular}.

Although we have highlighted the above games, the list is far from exhaustive. Others include: 
the \textbf{interpretability game}, where we pretend that the ML knows more than us and all we have to do is extract knowledge from it; \textbf{the causality game}, which tries to apply `billiard ball'-like causation to social systems; the \textbf{analytically tractable game} which assigns supposed extra value to a model with a mathematical solution; \textbf{the prediction game} which uses models to assign probabilities to future events; \textbf{the objective function game}, where we pretend that the reason we cannot solve a complex problem is because we do not yet know what it is trying to maximise; and Turing's \textbf{imitation game}, which equates general AI with the imitation of human conversation. As in our more detailed examples, all of these games reason about an open, complex system as if it were in fact a closed, model system.

\section{Conclusions}
\label{sec:conclusion}

By deconstructing ML games, 
we are not claiming they are never useful. 
Rather, we identify the points at which the games necessarily break down. Our aim is not to completely overturn ML practice, but rather to reveal that ML should be open and thus subject to changes not fully captured by any unique set of practices. The practices ML researchers adopt (or the games we play) are, at best, locally correct. Different, equally useful, practices can make contradictory suggestions. The arguments we have presented here are based on an assumption that the inherently open nature of complex systems means that they can never be captured by a single `objective' model. 
We have identified football as a concrete example in Section~\ref{sec:practice} and given a theoretical exposition of our position in Section~\ref{sec:what_is_complex}, and argued in Section~\ref{sec:games_we_play} 
that adopting our assumptions reveals extremely serious limitations and problems with the games played.

Revealing these limitations is particularly important because the field of AI currently garners unprecedented over-hype. The capabilities of models are repeatedly exaggerated, while uncritical trust in AI insulates the field and its claims from scrutiny. Currently, separating genuine concrete `progress' from deceptive vacuous over-hype can be challenging even to the most sensible and technically aware expert. The games played by ML researchers make the task of seeing progress clearly even more challenging. Instead of embracing Open ML, it is far too common to see researchers, in industry and academia alike,
emphasize the importance of their closed games. These researchers often appear oblivious to both the consequences of the games they are playing for the society we live in and to the limited assumptions upon which the rules of their games are designed.

While there is much to be gained by embracing an Open ML approach, there is even more to be lost from a failure to adopt it. When we build a single model of a complex adaptive system and present it as a true representation, we are not only committing a scientific fallacy, but also potentially causing harm. When models are 
integrated into decision making in the social world, the \textit{games} on which they are built can have dire downstream impacts. When these models fail (which they often do~\cite{birhane2021algorithmic,o2016weapons,noble2018algorithms,mehrabi2021survey}) people at the margins of society pay the heaviest price. Furthermore, harm and benefit are disproportionately distributed~\cite{birhane2021algorithmic}. Those that create and deploy ML models gain the most benefit and face the least harm. The negative consequences for those deploying these models are, at the very most, reptautional harm and/or critical scrutiny. Individuals and groups whom these models are applied to, on the other hand, benefit the least (if at all) and can potentially face dire situations such as death, imprisonment, loss of opportunities due to failure of or reductive models.   

We have placed our critique in the context of an approach, namely that of Open ML. We thus offer not just a critique (of great importance in and of itself) of existing practices, but also a way forward. 
We emphasize that capturing a complex system requires creating many different models at various levels and from different perspectives. 
The description of complexity in combination with the practical application 
provide both a context to the (more limited) approach based in theoretical games and a wider understanding of how progress can be made when applying ML in more complex domains.

This paper is by no means a comprehensive documentation of all the rich and varied ways in which ML is applied. Nor is it a complete description of the meta-games at play or a presentation of all the ethical and cultural consequences of seeing the world through these games. It is, though, an encouragement for those working in ML to recognise the limitations which modelling complex systems places on what we can and cannot achieve. In particular, many of the so-called big ideas in ML --- interpretability,  model-free learning, Bayesian updating of beliefs, general AI -–– are no more than games which describe our subjective practices. As such, many of the apparently `deep' results arising from these games are spurious illusions. They are tricks of the light shone on our own practices.

\bibliographystyle{vancouver}
\bibliography{sample}

\end{document}